\documentstyle[12pt]{article}

\newcommand{\be}{\begin{equation}}
\newcommand{\ee}{\end{equation}}
\newcommand{\bea}{\begin{eqnarray}}
\newcommand{\eea}{\end{eqnarray}}

\begin{document}
\begin{titlepage}

\vspace{1in}

\begin{center}
\Large
{\bf Tilted String Cosmologies}

\vspace{1in}

\large{Dominic Clancy$^{1a}$, Alexander Feinstein$^{2b}$, 
James E. Lidsey$^{3c}$\\
\& Reza Tavakol$^{1d}$}\\
\normalsize
\vspace{.4in}
$^1${\em Astronomy Unit, School of Mathematical Sciences,  \\
Queen Mary \& Westfield College, Mile End Road, London, E1 4NS, U.K.}\\
\vspace{.2in}
$^2${\em Dpto. de Fisica Teorica, Universidad del Pais Vasco, \\
Apdo. 644, E-48080, Bilbao, Spain}\\
\vspace{.2in}
$^3${\em Astronomy Centre and Centre for Theoretical Physics, \\
University of Sussex, Brighton, BN1 9QH, U.K.}
\end{center}
\vspace{.3in}
\baselineskip=24pt

\begin{abstract}
\noindent 
Global symmetries of the string effective action are employed to generate 
tilted, homogeneous Bianchi 
type ${\rm VI}_h$ string cosmologies from a previously known stiff perfect 
fluid solution to Einstein gravity. The dilaton 
field is not constant on the surfaces of homogeneity. 
The future asymptotic state of the models is 
interpreted as a plane wave and 
is itself an exact solution 
to the string equations of motion to all orders in the inverse string tension.
An inhomogeneous generalization of the Bianchi type III 
model is also found. 

\end{abstract}

PACS NUMBERS:  04.20.Jb, 04.50.+h, 11.25.Mj, 98.80.Cq   

\vspace{.3in}
$^a$Electronic address: dominic@maths.qmw.ac.uk

$^b$Electronic address: wtpfexxa@lg.ehu.es

$^c$Electronic address: jel@astr.cpes.susx.ac.uk

$^d$Electronic address: reza@maths.qmw.ac.uk

\end{titlepage}


All five perturbative string theories contain 
a Neveu--Schwarz/Neveu--Schwarz (NS--NS) sector  of bosonic 
fields, consisting of a graviton, $g_{\mu\nu}$, an antisymmetric 
two--form potential, $B_{\mu\nu}$, and a scalar dilaton field, 
$\phi$ \cite{gsw}. If we assume that the moduli 
fields arising from the internal dimensions are trivial, 
the effective, four--dimensional
action for these fields, to lowest--order in the 
inverse string tension, $\alpha'$, is  \cite{perry}
\begin{equation}
\label{effectiveaction}
S=\int d^4 x \sqrt{-g} e^{-\phi} \left[ R +\left( \nabla \phi \right)^2 
-\frac{1}{12} H_{\mu\nu\lambda}H^{\mu\nu\lambda} \right]  ,
\end{equation}
where $R$ is the Ricci curvature of the spacetime, 
$g \equiv {\rm det} g_{\mu\nu}$ and $H_{\mu\nu\lambda} 
\equiv \partial_{[\mu}
B_{\nu\lambda ]}$ is the field strength of the two--form. 
The one--loop $\beta$--function equations correspond to 
the field equations derived from this effective action. 

The dynamics of the very early universe below the string 
scale may have been determined by an action such as 
Eq. (\ref{effectiveaction}).
The general, spatially isotropic and homogeneous 
Friedmann--Robertson--Walker (FRW) string cosmology  
is known and well understood \cite{clw}. However, it is generally accepted 
that spatial anisotropies and inhomogeneities would have been important 
in the very early universe and the study of string cosmologies that 
relax the FRW assumptions is therefore well 
motivated. This is the purpose of the present work. 

The spatially homogeneous Bianchi \cite{bk,b} and Kantowski--Sachs \cite{ks}
cosmologies have been studied previously, 
and some inhomogeneous generalizations 
of these models were recently presented \cite{in}. 
Bianchi cosmologies admit 
a three--dimensional Lie group of isometries that acts simply--transitively 
on three--dimensional, spacelike orbits \cite{rs}.
The line element can be written in the form 
$ds^2 =-dt^2 +h_{ij}(t) \omega^i \omega^j$, where 
$h_{ij}$ is the metric on the surfaces of homogeneity, 
$t={\rm constant}$, and $\omega^i$ are one--forms that 
specify the Bianchi type. 
Barrow and Kunze have determined the general form of the 
field strength, $H_{\mu\nu\lambda}$, 
in these models when the two--form is spatially 
homogeneous, $B_{\mu\nu}=B_{\mu\nu}(t)$ \cite{bk}. Homogeneous 
solutions have also been found by 
Betakis and Kehagias, who imposed different restrictions on the components 
of the dual of the field strength \cite{b}. 

In all these previous studies, however, the dilaton field 
was assumed to be constant 
on the surfaces of homogeneity, $\phi =\phi (t)$. 
In this paper we find homogeneous Bianchi type ${\rm VI}_h$ models 
where the dilaton field is not constant on these surfaces. 
We refer to these models as {\em tilted} because the field
is spatially dependent. 
This can be consistent with the assumption of homogeneity if the pressure 
and energy density of the field remain homogeneous. Since these 
variables are defined in terms of the derivatives of the field, 
a linear dependence on one of the spatial coordinates is permitted. 

Type ${\rm VI}_h$ models are interesting because they have a non--zero 
measure in the space of homogeneous initial data \cite{bk}. 
The cosmologies  are 
generated by employing the global symmetries of 
the string effective action (\ref{effectiveaction}) \cite{t,t1,s}.  
A tilted, Bianchi type ${\rm VI}_h$, 
stiff perfect fluid solution
of Einstein gravity is taken as the seed solution \cite{w}. 
A given Bianchi model with  a perfect fluid source is
tilted if the fluid velocity vector is not
orthogonal to the spatial hypersurfaces \cite{ke}. Such a model appears 
spatially inhomogeneous to an observer comoving with the fluid, 
but homogeneous to one with a worldline that is orthogonal to the 
$t={\rm constant}$ surfaces. 

To proceed, we perform the conformal transformation:
\begin{equation}
\label{conformal}
\tilde{g}_{\mu\nu} = \Omega^2 g_{\mu\nu} , \qquad \Omega^2 
\equiv e^{-\phi}
\end{equation}
to the Einstein--frame, 
where the dilaton field is minimally coupled to gravity. 
The action (\ref{effectiveaction}) then takes the form 
\begin{equation}
\label{efe}
S=\int d^4 x \sqrt{-\tilde{g}} \left[ \tilde{R} 
-\frac{1}{2} \left( \tilde{\nabla} \phi \right)^2  -\frac{1}{12} 
e^{-2\phi} \tilde{H}_{\mu\nu\lambda}\tilde{H}^{\mu\nu\lambda} \right]
\end{equation}
and the metrics in the string-- and Einstein--frames 
are related by $ds^2_s =e^{\phi} ds^2_e$.
A stiff perfect fluid source for a homogeneous solution of Einstein gravity 
is dynamically equivalent to that of a massless, minimally coupled 
scalar field \cite{tt}. Eq. (\ref{efe}) therefore  
implies that such a solution may be reinterpreted as a 
truncated string cosmology with a 
vanishing two--form potential. 

In view of this, we 
consider the tilted, Bianchi type ${\rm VI}_h$ stiff 
perfect fluid cosmology
found by Wainwright, Ince and Marshman \cite{w}. The line element 
and fluid four--velocity are given by
\begin{equation}
\label{tilted}
ds^2_e =e^f \left( -dt^2 + dz^2 \right) +e^{-2z} \sinh 2t 
\left( e^p dx^2 +e^{-p} 
dy^2 \right)  
\end{equation}
and 
\begin{equation}
\label{velocity}
u_{\mu} =\frac{\varphi_{, \mu}}{
\left| \varphi_{, \nu} \varphi^{, \nu} \right|^{1/2}}   ,
\end{equation}
respectively, where
\begin{eqnarray}
\label{h}
f = \frac{1}{2} \left( k^2 +m^2 +\alpha^2 +\beta^2 -1 \right) \ln \sinh 2t +
(km +\alpha \beta ) \ln \tanh t \\
\label{p}
p=-2kz +k\ln \sinh 2t +m \ln \tanh t 
\end{eqnarray}
represent the longitudinal and transverse components of the gravitational 
field, the scalar quantity, $\varphi$, is given by  
\begin{equation}
\label{varphi}
\varphi = -2\alpha z +\alpha \ln \sinh 2t +\beta \ln \tanh t   ,
\end{equation}
the constants $\{ \alpha , \beta , k,m \}$ 
satisfy the constraint
\begin{equation}
\label{constraint}
\beta^2 -\alpha^2 +m^2 -k^2 -3 =0 
\end{equation}
and the constant $k \equiv (-h)^{-1/2}$ is related to the group parameter, 
$h<0$. 

The fluid flow is not orthogonal to the group orbits if $\alpha \ne 0$ and 
the cosmology is in general tilted \cite{w}. It is 
orthogonal when $\alpha =0$ and reduces to the Ellis--MacCallum vacuum 
solution when $\alpha =\beta =0$ \cite{em}. The cases $k^2 =1$ and $k=0$ 
represent tilted Bianchi type III and V models, respectively. 
The solution (\ref{tilted})--(\ref{constraint}) 
may be viewed as a truncated string cosmology with a trivial two--form 
potential if the dilaton field is identified with $\varphi$ in 
Eq. (\ref{varphi}) and if Eq. (\ref{tilted}) is interpreted 
as the Einstein--frame metric. 

The transverse space is spanned by the variables $x^a=(x,y)$ and we assume 
that it has a toroidal topology, $S^1 \times S^1$. The components 
of the metric (\ref{tilted}) are independent 
of $x^a$ and it therefore 
admits an abelian group of isometries, $G_2$ \cite{c}. This group  
acts on the two--dimensional orbits and the two, commuting spacelike 
Killing vectors are $\partial /\partial x^a$. When all 
massless degrees of freedom in Eq. (\ref{effectiveaction}) 
are independent of $x^a$, the field equations 
exhibit a global ${\rm O}(2,2)$ symmetry \cite{t}. 
Application of this symmetry to a given $G_2$ solution 
leads to a new, inequivalent solution. 

The ${\rm O}(2,2)$ symmetry becomes manifest in the string 
frame when the only non--trivial component of the two--form 
potential is $B_{xy}=B_{xy} (t,z)$ \cite{t}. 
We now employ the tilted, type ${\rm VI}_h$ 
cosmology (\ref{tilted})--(\ref{constraint}) as a seed for 
generating a class of string cosmologies with a non--trivial two--form. 
The isomorphism ${\rm O}(2,2) = {\rm SL}(2,R)
\times {\rm SL}(2,R)$ implies that an ${\rm O}(2,2)$
symmetry transformation may be expressed in terms of 
two ${\rm SL}(2,R)$ transformations \cite{t1,i}. One 
of these corresponds to an Ehlers transformation 
and leaves the two--form potential 
invariant \cite{ehlers,g}. The other may be employed to 
generate a non--trivial two--form potential from a solution 
where such a field is zero. 
If we denote the 
string--frame metric on the transverse space as $\Gamma_{ab}$, 
with determinant $\Gamma \equiv {\rm det}\Gamma_{ab}$, and define a 
complex scalar field $\rho \equiv B_{xy} +i \sqrt{\Gamma}$, 
this ${\rm SL}(2,R)$ transformation is given by \cite{i}
\begin{equation}
\label{barrho}
\bar{\rho}=\frac{a\rho +b}{c\rho +d} ,
\end{equation}
where $ad-bc =1$. 

Thus, for the case where $B_{xy} =0$, 
Eq. (\ref{barrho}) implies that 
\begin{eqnarray}
\label{barGamma}
\bar{\Gamma} =\frac{\Gamma}{(d^2 +c^2 \Gamma )^2} \\
\label{barB}
\bar{B}_{xy} =\frac{ac\Gamma +bd}{c^2 \Gamma +d^2}  , 
\end{eqnarray}
where 
\begin{equation}
\label{Gammaform}
\Gamma = e^{2\phi -4z} \sinh^2 2t
\end{equation}
and $\phi$ is given by Eq. (\ref{varphi}). 

The `shifted' dilaton field, $\Phi \equiv \phi -(\ln \Gamma 
)/2$, and the longitudinal component of the string--frame metric
are singlets under Eq. (\ref{barrho}) \cite{t,t1}. This 
implies that 
\begin{eqnarray}
\label{barphi}
e^{\bar{\phi}} =\frac{e^{\phi}}{d^2 +c^2 \Gamma} \\
\label{barh}
\bar{f} =f +\phi -\bar{\phi}   ,
\end{eqnarray}
where $f$ is given by Eq. (\ref{h}). 
The components of the transverse 
string--frame metric transform as 
$\bar{\Gamma}_{aa} = \Gamma_{aa} ( \bar{\Gamma}/\Gamma )^{1/2}$ and 
$\bar{\Gamma}_{xy} =0$ 
\cite{i}. The corresponding metrics in the string-- and Einstein--frames 
are therefore given by 
\begin{equation}
\label{stringdual}
d\bar{s}^2_s =e^{f+\phi} \left( -dt^2 +
dz^2 \right) +\frac{\Gamma^{1/2}}{d^2 +c^2\Gamma} 
\left( e^p dx^2 +e^{-p} dy^2 \right) 
\end{equation}
and 
\begin{equation}
\label{einsteindual}
d\bar{s}^2_e =e^f \left( d^2 +c^2 \Gamma \right)
\left( -dt^2 +dz^2 \right) + \Gamma^{1/2} e^{-\phi} 
\left( e^p dx^2 +e^{-p} dy^2 \right)    ,
\end{equation}
respectively. 
The transverse part of the 
Einstein--frame metric is invariant under the transformation 
(\ref{barrho}). 

In general, the dual solution (\ref{einsteindual}) 
does not preserve the one--forms of the Bianchi type ${\rm VI}_h$ 
metric. Two of the one--forms have a direct dependence 
on the spatial variable $z$ and this leads to the linear dependence 
exhibited by the transverse component, $p$, in Eq. (\ref{p}) \cite{rs}. 
The third one--form is $dz$ and can not be multiplied 
by a function of $z$ if the spacetime is to remain 
homogeneous. Since the longitudinal 
component of the metric (\ref{einsteindual}) is 
$z$--dependent, homogeneity is broken along this direction 
and the solution therefore represents an inhomogeneous $G_2$ cosmology 
\cite{c}. 

On the other hand, the determinant $\Gamma$ 
is independent of the spatial coordinates
if $\alpha =-1$, i.e., $\Gamma = \tanh^{2\beta} t$. In this case, 
the Einstein--frame metric is an homogeneous type 
${\rm VI}_h$ spacetime. The two--form potential (\ref{barB}) 
is spatially homogeneous 
for this value of $\alpha$, but the model is tilted 
due to the spatial dependence of the dilaton field (\ref{barphi}). 
The string--frame metric (\ref{stringdual}) is not 
homogeneous, however, because the longitudinal component of the metric 
acquires a $z$--dependence from the dilaton. 

The string effective action (\ref{effectiveaction}) exhibits a further
global ${\rm SL}(2, R)$ symmetry that may 
be employed to generate a second class of models from 
the solution (\ref{tilted})--(\ref{constraint}) \cite{s}. 
The symmetry is manifest in the Einstein--frame 
when formulated in terms of the dual 
of the field strength, $\tilde{H}^{\mu\nu\lambda} \equiv  e^{2\phi} 
\tilde{\epsilon}^{\mu\nu\lambda\kappa} \tilde{\nabla}_{\kappa} 
\sigma$, where $\tilde{\epsilon}^{\mu\nu\lambda\kappa}$ is the 
covariantly constant four--form and $\sigma$ is interpreted as a 
pseudo--scalar axion field . The dilaton and axion fields 
then parametrize the ${\rm SL}(2,R)/{\rm U}(1)$ coset. The 
${\rm SL}(2,R)$ transformation acts on the  
complex scalar field $\chi \equiv \sigma +ie^{-\phi}$ such that 
$\bar{\chi} = (a'\chi +b')/(c' \chi +d')$, where $a'd' -b'c' =1$ 
\cite{s}. 

When the axion field is initially zero, 
the transformed dilaton and axion fields are therefore given by
\begin{eqnarray}
\label{transdilaton}
e^{\bar{\phi}} = c'^2 e^{-\phi} +d'^2 e^{\phi} \\
\label{tsigma}
\bar{\sigma} =\frac{a'c'e^{-\phi} +b'd'e^{\phi}}{c'^2e^{-\phi} 
+d'^2 e^{\phi}}    ,
\end{eqnarray}
where $\phi$ is given by Eq. (\ref{varphi}). The Einstein--frame metric is 
invariant under this ${\rm SL}(2, R)$ transformation and 
the string--frame metric is given by \cite{s}
\begin{equation}
\label{Sdual}
d\bar{s}^2_s =
e^{\bar{\phi}} ds^2_e     ,
\end{equation}
where $ds^2_e$ is given by Eq. (\ref{tilted}). The $\alpha =-1$ solution 
generated by 
this ${\rm SL}(2,R)$ transformation is different to 
Eq. (\ref{stringdual}), since the longitudinal 
part of the string--frame metric is not 
invariant under Eq. (\ref{Sdual}). Moreover, the two--form 
potential (\ref{barB}) derived from the 
transformation (\ref{barrho})
is spatially independent and this 
implies that the axion field must be 
independent of time. For the solution (\ref{tsigma}), 
however, the axion depends on both space and time. 

The late--time asymptotic behaviour of both 
solutions (\ref{stringdual}) and (\ref{Sdual})
may be determined in the Einstein--frame. For the solution 
(\ref{stringdual}) with $\alpha =-1$, 
$\Gamma \rightarrow 1$ as $t \rightarrow \infty$ 
and Eq. (\ref{barB}) then implies that the two--form
potential approaches a constant value. Moreover, Eq. (\ref{barphi}) 
implies that the dilaton field asymptotes to 
its original form, modulo a constant shift in its value, and 
we may specify $d^2 +c^2 \Gamma =1$ without loss of generality. 
The original seed solution (\ref{tilted}) is 
therefore recovered in this late--time limit. 

For the solution (\ref{Sdual}), it follows from Eq. (\ref{transdilaton}) that 
$\bar{\phi} \rightarrow \infty$. The axion field, $\sigma$, approaches 
a constant value in this limit and becomes 
dynamically negligible. In both models, therefore, 
the solutions asymptote to the seed solution (\ref{tilted}). 
The late--time limit of this metric when $\alpha =-1$ can be analyzed 
by defining new coordinates 
\begin{eqnarray}
\label{u}
u \equiv \frac{1}{2+k^2} \exp\left[ (2+k^2)(t-z) \right] \\
\label{v}
v \equiv \frac{2^{-(2+k^2)}}{2+k^2} \exp \left[ 
(2+k^2)(t+z) \right]
\end{eqnarray}
and rescaling the transverse coordinates. 
The line element  (\ref{tilted}) then tends to the Rosen form 
of a plane wave:
\begin{equation}
\label{uvmetric}
d s^2_e =-dudv +u^{2(1+k)/(2+k^2)} dx^2 + u^{2(1-k)/(2+k^2)}
dy^2
\end{equation}
and the dilaton field becomes null:  
\begin{equation}
\label{uvphi}
\phi =\phi_0 -\frac{2}{2+k^2} \ln u    ,
\end{equation}
where $\phi_0$ is a constant. 

Defining a further set of variables 
$\{ \bar{v}, \bar{x} , \bar{y} \}$:
\begin{eqnarray}
\label{newv}
v \equiv \bar{v} -\frac{1}{(2+k^2)u} \left( \bar{x}^2 +\bar{y}^2 +2k 
\bar{x}\bar{y} \right) \\
x \equiv \frac{1}{\sqrt{2}} u^{-(1+k)/(2+k^2)} \left( \bar{x} 
+ \bar{y} \right) \\
y \equiv \frac{1}{\sqrt{2}} u^{-(1-k)/(2+k^2)} 
\left( \bar{x} -\bar{y} \right)
\end{eqnarray}
implies that the metric (\ref{uvmetric}) may also be written 
as a plane wave in Brinkmann coordinates \cite{g}:  
\begin{equation}
\label{newuvmetric}
ds_e^2 =-dud\bar{v} +F(u, \bar{x} , \bar{y} )du^2 
+d\bar{x}^2 +d\bar{y}^2    ,
\end{equation}
where
\begin{equation}
\label{F}
F=-\frac{1}{(2+k^2)^2} \left[ \bar{x}^2 +\bar{y}^2 +2k^3 
\bar{x}\bar{y} \right] \frac{1}{u^2}
\end{equation}

The dilaton field (\ref{uvphi}) and metric (\ref{newuvmetric}) 
are independent of the variable 
$\bar{v}$ and the solution 
admits a covariantly constant, null Killing 
vector field, $l^{\mu}$, such that 
$l_{\mu}l^{\mu} = \nabla_{\mu} 
l_{\nu} =0$. 
Since the gradient of the dilaton field 
is proportional to $l_{\mu}$, 
the one--loop $\beta$--function equations reduce to the 
single constraint $\tilde{R}_{uu} = (\partial_u \phi )^2/2$ when 
expressed in terms of the Einstein--frame variables. It can 
be verified by direct substitution 
that this constraint is identically 
satisfied. Thus, the asymptotic form of the tilted solutions
in the late time limit is itself an exact 
solution to the one--loop $\beta$--function equations. 

Moreover, all higher--order terms in the $\sigma$--model perturbation 
theory are second--rank tensors constructed 
from powers of the Riemann tensor, the gradient of the dilaton 
field, the metric, the three--form field strength and their 
covariant derivatives \cite{gsw}. 
It can be shown by applying a geometrical argument 
due to Horowitz and Steif \cite{hs} that a plane--wave background of the 
form (\ref{uvphi}) and (\ref{newuvmetric}) acquires no 
$\alpha'$--corrections 
because the Riemann tensor is orthogonal to  $l^{\mu}$ and $\nabla^{\mu}$
on all its indices and because $\nabla_{\mu} \phi$ 
is proportional to  $l_{\mu}$. 
Consequently, the asymptotic solution (\ref{uvphi}) 
and (\ref{newuvmetric}) represents an {\em exact} 
solution to the classical string equations of motion to 
{\em all} orders in the inverse string tension. 

The asymptotic form of the type III $(h=-1)$ cosmologies  
is of particular interest. For the tilted solution generated 
by the ${\rm SL}(2,R)$ transformation (\ref{barrho}), 
the string--frame metric 
asymptotes to $d\bar{s}^2_s =-dUdV  +U^2 dx^2 +U^{-2} dy^2$, where 
$U\equiv u^{1/3}$, $V\equiv 3v$ and we have specified  $k=1$ 
without loss of generality. 
Since there exists an abelian 
isometry $\tilde{y}= y +\gamma$ for some constant $\gamma$ in this metric, 
we may consider a Buscher transformation associated 
with this transverse coordinate \cite{bu}. For the case where
$ac=-bd$,  
the two--form potential asymptotically vanishes
and the Buscher transformation generates an 
inequivalent solution, where $\bar{\Gamma}_{yy} = \Gamma_{yy}^{-1}$, 
$\bar{\phi} = \phi -\ln \Gamma_{yy}$ and  
all other components remain invariant. Thus, 
the tilted type III model with $ac=-bd$ is dual to 
$ds_s^2 =-dUdV +U^2(dx^2+dy^2)$ with a constant dilaton 
and this is the Milne form of flat space expressed in an expanding frame. 

To summarize thus far, we have found homogeneous models that can
be interpreted as string cosmologies, where the dilaton field 
is not constant on 
the surfaces of homogeneity. The models are homogeneous in the conformally 
related Einstein--frame and asymptote to a 
perturbatively exact plane wave at late times. It would be 
interesting to show that this plane wave is also an 
exact solution at the non--perturbative level, i.e., that 
it corresponds to an exact conformal field theory (CFT). 
This is important because the propagation of first--quantized 
strings in a particular background 
can be determined by interpreting the string solution as a two--dimensional 
CFT. (For a review see, e.g., Ref. \cite{tsey}). Amati 
and Klimcik have shown that Ricci flat 
plane wave solutions are non--perturbatively exact \cite{amati}. 
The question that arises is whether their method can be extended 
to include a non--trivial dilaton field. 

Another interesting limit is associated with the asymptotic state of 
the inhomogeneous $G_2$ model (\ref{stringdual}) generated by 
the transformation (\ref{barrho}) from the orthogonal type 
III stiff perfect fluid cosmology, where $\alpha =0$ and $k=1$. 
We specialise to the case where $a=0$ and $b=-c=1$ 
and assume that $d$ is arbitrary. The string--frame metric 
(\ref{stringdual}), dilaton field (\ref{barphi}) and two--form 
potential (\ref{barB}) are then given by 
\begin{eqnarray}
\label{cft}
d\bar{s}^2_s =-dwdr +\frac{w}{d^2+w^2} \left( wdx^2 +w^{-1} dy^2 \right) \\
\label{cftphi}
\bar{\phi}=-\ln (d^2 +w^2 ) \\
\label{cftB}
\bar{B}_{xy} =\frac{d}{d^2+w^2}
\end{eqnarray}
in the limit $t\rightarrow \infty$, where $w \equiv e^{2(t-z)}/2$ 
and $r\equiv e^{2(t+z)}/8$. 

In the limit $d=0$, 
the metric (\ref{cft})  is related to the Bianchi III 
form of flat space, $ds^2=-dwdr +dx^2+ w^2 dy^2$, by a 
Buscher transformation. The existence of a covariantly constant 
null Killing vector field implies that the solution 
(\ref{cft})--(\ref{cftB}) acquires no higher--order 
corrections and is therefore perturbatively exact to all orders. 
However, it also admits a CFT interpretation
in terms of a gauged Wess--Zumino--Witten (WZW) model
\cite{tsey,st,rev}. WZW models correspond to conformally invariant 
non--linear $\sigma$--models. 
The background (\ref{cft})--(\ref{cftB}) can be derived by taking 
a special singular limit of the gauged $[{\rm SU}(2) \times {\rm SL}(2,R)]/
[{\rm U}(1) \times {\rm R}]$ WZW model \cite{st}\footnote{The reader is 
referred to Ref. \cite{tsey} for a review of WZW models
and the derivation of this result.}. 

The solutions presented above are also relevant 
to the pre--big bang inflationary scenario
\cite{pbb}. The 
fundamental postulate of this scenario is that the initial 
state of the universe should be in the deeply perturbative regime of 
small curvature and coupling. Inflation is then driven by the kinetic 
energy of the dilaton field and this causes the curvature and 
coupling to grow. The Einstein--frame metric (\ref{tilted})
asymptotes to a spatially flat,  Bianchi type I form
at early times, 
$t \rightarrow 0$. The solution in this limit may be interpreted as a 
generalization of the Kasner metric to include a massless scalar field, 
where each of the three scale factors vary as some power of 
synchronous time. For  a certain 
region of parameter space, 
the time--reversal of 
a solution of this form  corresponds to a pre--big bang, 
superinflationary expansion in the string frame \cite{clt}. 

A generic initial state for the pre--big bang cosmology has 
recently been formulated in terms of an arbitrary 
mixture of incoming gravitational and scalar waves
\cite{bdv}. The time
symmetry of the cosmological field equations implies that 
the future asymptotic states of the models considered in this 
work can be reinterpreted as past initial states for pre--big 
bang cosmologies. The plane wave 
solutions (\ref{newuvmetric}) and (\ref{cft}) 
provide analytical examples of an initial state 
for a universe that evolves into a
Kasner--like behaviour. It would be interesting to consider 
such a transition further 
and, in particular, to study the role played by the tilt of the dilaton 
field in the pre--big bang scenario. 

\vspace{.3in}

D. C. was supported by the Particle Physics 
and Astronomy Research Council, PPARC, 
A.F. was supported by Spanish Science Ministry Grant
172.310-0250/96 and Basque Country University Grant UPV 172.310-EB150/98, 
J. E. L. was supported by the Royal Society and R. T. benefited 
from PPARC UK Grant No. L39094.

\end{document}